\documentclass[conference]{IEEEtran}

\usepackage{blindtext, graphicx}
\usepackage{cite}
\usepackage{float}
\usepackage{graphicx}
\usepackage[justification=centering]{caption}
\usepackage{caption}
\usepackage{refstyle}
\usepackage{url}
\usepackage{mathtools}
\usepackage[flushleft]{threeparttable} % http://ctan.org/pkg/threeparttable
\usepackage{booktabs,caption}
\usepackage{amsmath}
\usepackage[hidelinks,bookmarks=false]{hyperref}
\usepackage[table]{xcolor}
\usepackage{pdflscape}
\usepackage{pdflscape}
\usepackage{afterpage}
\usepackage{capt-of}	% or use the larger `caption` package
\usepackage[table]{xcolor}
\usepackage{colortbl}
\usepackage[table]{xcolor}
\usepackage{tabularx,booktabs} 
\usepackage{caption}
\usepackage{multirow}
\usepackage{multicol}
\usepackage{float}
\usepackage{algorithm}
\usepackage{algpseudocode}
\usepackage{lipsum,multicol}
\usepackage{balance}

\makeatletter
\def\BState{\State\hskip-\ALG@thistlm}
\makeatother
\usepackage{array, threeparttable} % to add footnotes to the tables
\usepackage{authblk}

\ifCLASSINFOpdf
\else
\fi

\begin{document}
\title{On the Energy Consumption Forecasting \\ of Data Centers Based on Weather Conditions:\\ Remote Sensing and Machine Learning Approach}

\author{Georgios Smpokos$^1$$^,$$^3$, Mohamed A. Elshatshat$^2$$^,$$^4$, Athanasios Lioumpas$^1$ and Ilias Iliopoulos$^1$
	
	\\
	\small{$^1$ Division of Technology Management \& Wholesale Market - CYTA Hellas, Greece}\\
	\small{$^2$ Institute of Computer Science - Foundation for Research and Technology - Hellas (ICS-FORTH), Greece}\\
	\small{$^3$Department of Science \& Technology - Link\"{o}ping University, Sweden}\\
	\small{$^4$Computer Science Department - University of Crete, Greece}\\
	\small{emails:~\{georgios.sbokos,~athanasios.lioumpas,~ilias.iliopoulos\}@hq.cyta.gr, mohamed@ics.forth.gr}
}


\maketitle

\begin{abstract}
%\boldmath

The energy consumption of Data Centers (DCs) is a very important figure for the telecommunications operators, not only in terms of cost, but also in terms of operational reliability. A reliable weather forecast would result in a more efficient management of the available energy and would make it easier to take advantage of the modern types of power-grid based on renewable energy resources. In this paper, we exploit the capabilities provided by the FIESTA-IoT platform in order to investigate the correlation between the weather conditions and the energy consumption in DCs. Then, by using multi-variable linear regression process we model this correlation between the energy consumption and the dominant weather condition parameters in order to effectively forecast the energy consumption based on the weather forecast. This procedure could be part of a wider resources optimization process in the core network towards an end-to-end (e2e) access/core network optimization of resources utilization. We have validated our results through live measurements from the RealDC testbed. Results from our proposed approach indicate that forecasting of energy consumption based on weather conditions could help not only DC operators in managing their cooling systems and power usage, but also electricity companies in optimizing their power distribution systems.

\end{abstract}


\IEEEpeerreviewmaketitle

\section{Introduction}
Data centers (DCs) are scaled up as the IT sector evolves rapidly with increasing demand in data throughput and cloud storage capacities (caching). Among the crucial aspects that DC operators have to deal with is the energy consumption of both the IT infrastructure and their cooling systems. The a priori knowledge of the expected consumption is very important not only in terms of cost but also in terms of operational continuity and reliability in the DC business. As a large portion of the power consumption is used by the cooling systems, in this study we aim to investigate the relation between the weather conditions and the DC power consumption which will indicate that weather data could be used for forecast models for energy consumption. A reliable forecast of this consumption would result in a more efficient management of the available resources and would make it easier to take advantage of the modern types of power distribution systems based on renewable energy sources.

\begin{figure}[t!]
	\centering
	\includegraphics[width=0.4\textwidth]{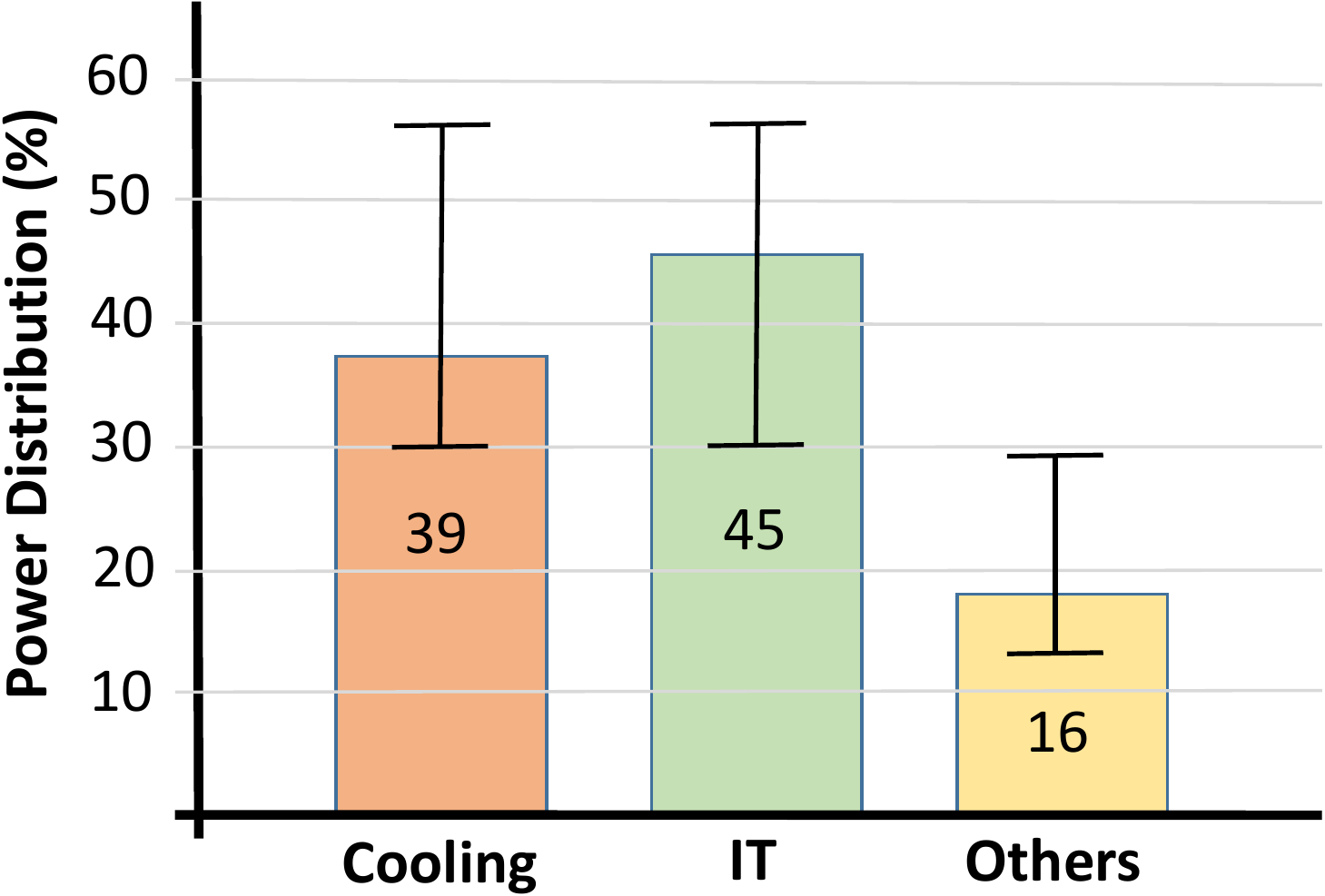}
	\caption{Power distribution of a typical data center.}
	\label{fig:figure1.pdf}
	
\end{figure}

As a large portion of the energy consumption of DCs is driven towards cooling the IT infrastructure, it is of great interest to investigate the factors that affect it \cite{7279063}. To this end, this study investigates how the outdoor weather conditions are related to the energy consumption of the DC. The IT hardware equipment residing in the DC creates the demand for the power and cooling. As shown in Fig. \ref{fig:figure1.pdf}, cooling the DC requires a significant amount of power and as such represents a significant factor to improve its efficiency \cite{song2015, 3GPP2014}. In DCs, about 40\% of the total energy is consumed for cooling the IT equipment. Cooling costs are one of the major contributors to the total operational expenditure (OPEX) of large DCs. There is a variety of energy efficiency metrics for DC cooling, such as Power Usage Effectiveness (PUE), Coefficient of Performance (COP) and chiller hours. The results of \cite{song2015} show that the cooling efficiency and operating costs vary significantly with different climate conditions, energy prices and cooling technologies. As weather conditions is a major factor which affects the ventilation and air cooling systems, it is of great significance to study their correlation and deploy efficient energy management techniques. For the modeling of the energy consumption of data centers we refer the reader to the comprehensive survey in \cite{7279063} and the references therein.

This study presents the results of the field experiment titled DC-IoT and implemented on the FIESTA-IoT platform using the RealDC testbed sensors data \cite{3GPP2014}. FIESTA-IoT platform provided a blueprint experimental infrastructure for heterogeneous IoT technologies and it was an appropriate enabler towards the realization of this experiment as it offered:
\begin{itemize}
	\item the capability to fetch IoT data from different testbeds based on a variety of filtering options, i.e., location, type of measurement, timestamp filtering etc.
	\item an interface that can interwork with open-source software and platforms (e.g., REST, SPARQL), which enables the easy and straightforward implementation of algorithms and procedures.
	\item an almost real-time ticketing and supporting platform that accelerates the implementation of the experiments and optimizes how the procedures are applied in practice.\\
\end{itemize}

The contribution of this study is summarized as follows:
\begin{itemize}
	\item to study and analyze the correlation between the weather conditions and the energy consumption of data centers using the historical data of the RealDC testbed. The output of this analysis is used to produce a statistical model that correlates the weather variables with the energy consumption.
	\item to exploit the statistical model to construct a forecast model that is able to predict the energy consumption of the data center with regards to the weather forecast. The model is verified through comparison with the real resources consumption measurements from the RealDC testbed.
	\item to provide an energy consumption forecast procedure to stakeholders and data center operators with estimates of their resource utilizations that enables to optimize their power distribution and energy consumption depending on the weather conditions. 
\end{itemize}

\section{Background and Related Work}

\begin{figure}[t!]
	\centering
	\includegraphics[width=0.5\textwidth]{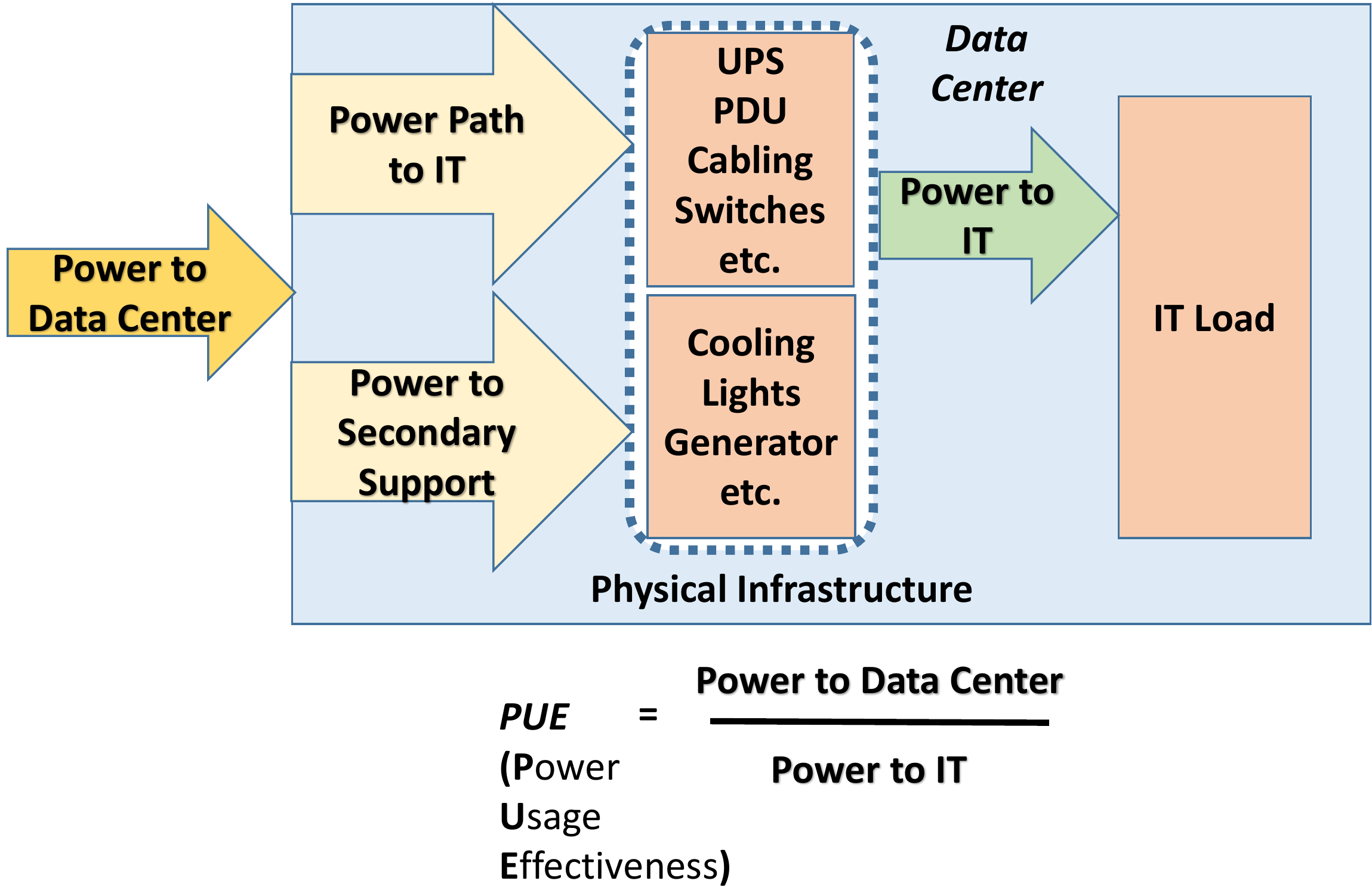}
	\caption{Power flows for a data center and power usage effectiveness calculation.}
	\label{fig:figure2.pdf}
\end{figure}

In Fig. \ref{fig:figure2.pdf} we provide an overview of the power allocation and processes of a data center. Climate and weather conditions affect the Power Usage Effectiveness (PUE), that is why it tends to be lower during cooler periods of the year according to \cite{7279063}. The DC operator's goal is to maintain a low PUE value \cite{google} thus spending less on cooling of the DC IT equipment. There are many solutions for DC energy efficiency management in areas with extreme climate conditions as illustrated in \cite{Cho2012}. The total operational cost depends on the type of air quality control systems such as air conditioning, humidifiers and water chillers. Thus, to reduce this operational cost many DC operators now build their DCs in cooler climate areas as in northern Europe \cite{bbc}. Consequently, the temperature conditions of the DC location is an important parameter that could reduce cost \cite{Cho2012,Berezovskaya2016}. According to \cite{abb2015}, by using renewable energy for power consumption within a DC environment, the total operating expenses could be decreased. Many DC owners are deploying renewable energy generation plants next to their DC so weather conditions is of great interest in the forecast and production of energy making their system more autonomous, more reliable and cost efficient.

The majority of air cooling systems are based on importing air from the DC environment. Other solutions include liquid cooling techniques where water is introduced in the cooling facilities in order to export heat from the IT infrastructure. Liquid cooling has been applied lately in many cases, with liquids being more efficient at expelling heat than air. Evaporative cooling is another energy efficient technique, especially applicable in dry climates. However, evaporative cooling often sparks a debate over the use of additional water, especially in water-constrained areas. Water usage is a topic that is gaining increased attention and will continue to do so in the near future \cite{ron2012}. All of the air and liquid cooling systems increase the energy consumption of the DC and make it very important for operators to optimize the usage of water resources as well as using free cooling (no additional cooling mechanism-direct external air introduction) when the operational temperature restrictions are not tight (higher acceptable IT hardware operational temperature). Nevertheless, there are no activities regarding the exploitation of the expected weather conditions towards adapting the operational parameters of cooling systems accordingly.  

With the recent advances in the applications of machine learning for prediction models, few studies have considered exploiting these models in forecasting the energy consumption of data centers \cite{7764178, 6313813, 7421894, AHMAD2014102}. In \cite{7764178} and \cite{6313813}, forecasting of the generic energy consumption within different urban locations was carried out using support vector machines (SVMs) and artificial neural networks (ANNs) algorithms. Both studies investigated the influence of weather conditions on the energy consumption and the results showed the effectiveness of prediction models based on weather conditions in forecasting. In \cite{7421894}, ANN algorithms  were used to forecast the energy consumption of a cloud computing system based on experimental data from a testbed. Finally, the authors in \cite{AHMAD2014102} provided a review of energy consumption forecasting methods with specific emphasis on ANN algorithms to forecast the energy consumption in residential buildings. However, based on the referenced work, none of these have exploited the effects of weather conditions to forecast the energy consumption of data centers.

\section{Experiment Setup}
\begin{figure}[h!]
	\centering
	\includegraphics[width=0.48\textwidth]{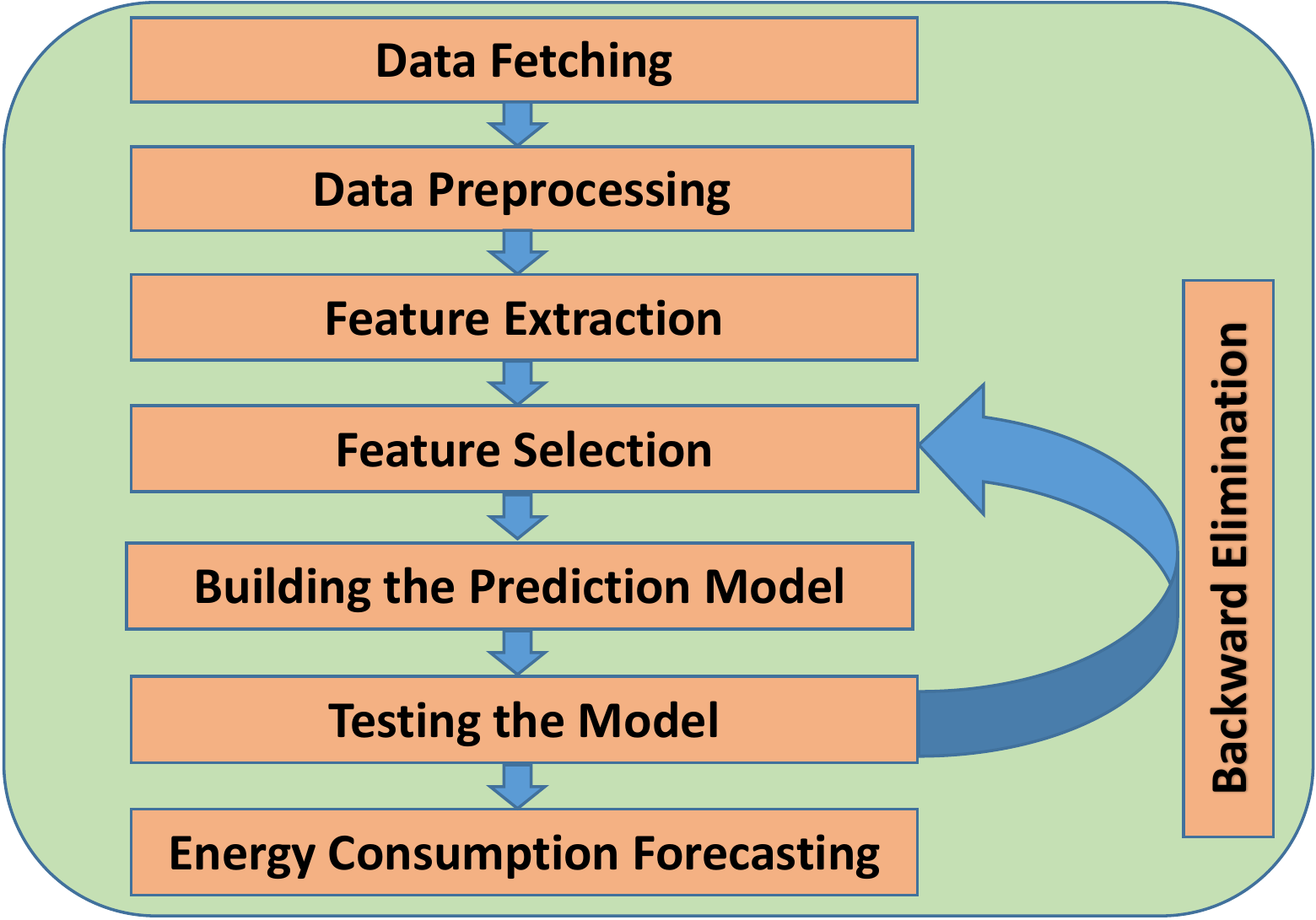}
	\caption{Experiment steps.}
	\label{fig:figure5.pdf}
	\vspace{0.4cm}
\end{figure}

In this section, we describe the detailed implementation of the experiment steps as shown in Fig. \ref{fig:figure5.pdf}. The main objective of the DC-IoT experiment is to correlate the energy consumption with the weather conditions near the DC in order to create a forecast model that could predict accurately the energy consumption (active and reactive power). The goal of this forecast model is to predict the power requirements of a DC with respect to the future weather conditions. By using the weather and the energy consumption data from the DC, the forecast model is trained and used for future predictions.\\

\subsection{Data Fetching and Preprocessing}
\begin{figure}[t!]
	\centering
	\includegraphics[width=0.49\textwidth]{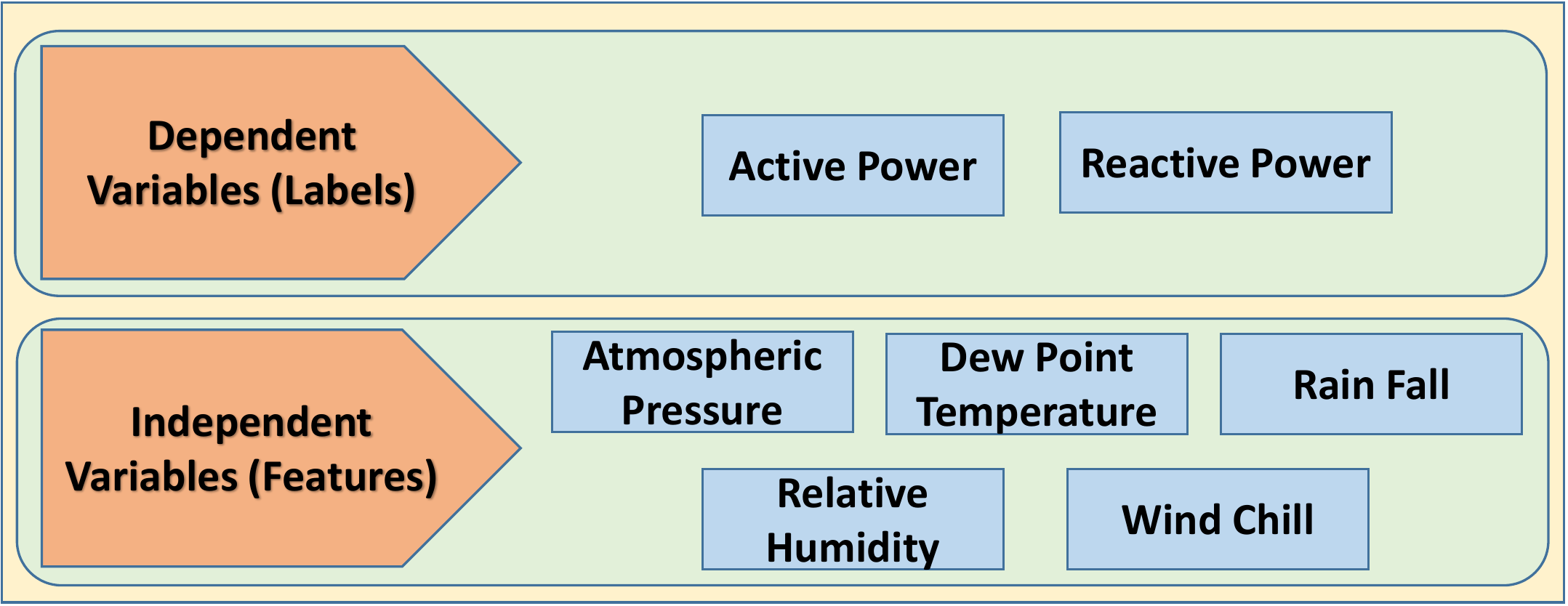}
	\caption{Dependent and independent variables.}
	\label{fig:figure6.pdf}
	\vspace{0.4cm}
\end{figure}

The first step in our experiment was to fetch data related to the weather conditions from the RealDC testbed sensors. The process of fetching the data was performed in a synchronous manner to ensure that the accuracy of the historical weather measurements could be used to build a reliable forecasting model. The fetching was done through SPARQL queries and python scripts, that implemented the iterations of the queries and the storage of the data to local files and databases for the off-line preprocessing. The data acquisition requests were implemented through a python script to automate most of the request-fetching procedures and store the data in comma separated value (csv) files. Afterwards, data pre-processing was essential in order to be able to extract the most dominant weather condition features that were correlated with the energy consumption. These features include the dew point temperature, the wind chill levels, the rainfall data, the atmospheric pressure and the relative humidity and are used for building the forecast model. The energy consumption of the DC is reflected in the active and reactive power measurements. Fig. \ref{fig:figure6.pdf} shows all the dependent and independent variables fetched from the testbed.

One essential step in the pre-processing phase was to ensure that all the data were synchronized by forming timestamp ordered data structures including the independent variable data and the active/reactive power values of each available power sensor. The timestamp of a few power measurement observations was not aligned (not synchronized - 5 minutes shift) compared to the weather measurements timestamp. For this reason we had to shift the timestamp of the power measurements that were not synchronized for a small time value (5 minutes). Additionally, we filtered out any measurements that did not include all the desired data (i.e., weather and energy) for the same timestamp. Eventually, we managed to create csv formatted files that included in each of their rows the weather data plus the active/reactive power values from each sensor. \\

\subsection{Feature Extraction and Selection}
Feature extraction indicates the process of representing the weather conditions parameters obtained in the previous step in the form of features that can be used in the prediction step to build the regression model. Additionally, feature selection is a recursive step that is performed to ensure that only the most dominant features are used to predict the energy consumption. For our analysis, we had to find out which power consumption measurements were more correlated to the weather independent variables. The correlation of the weather data and the power consumption (active-reactive power) was implemented using R programming utilizing its statistics library resources. By evaluating the correlation of active/reactive power values with weather sensor data we managed to find which sensor (active-reactive power) was more correlated to the weather conditions.

\section{Prediction Model \\(Supervised Learning Approach)}

\begin{figure}[t!]
	\centering
	\includegraphics[width=0.5\textwidth]{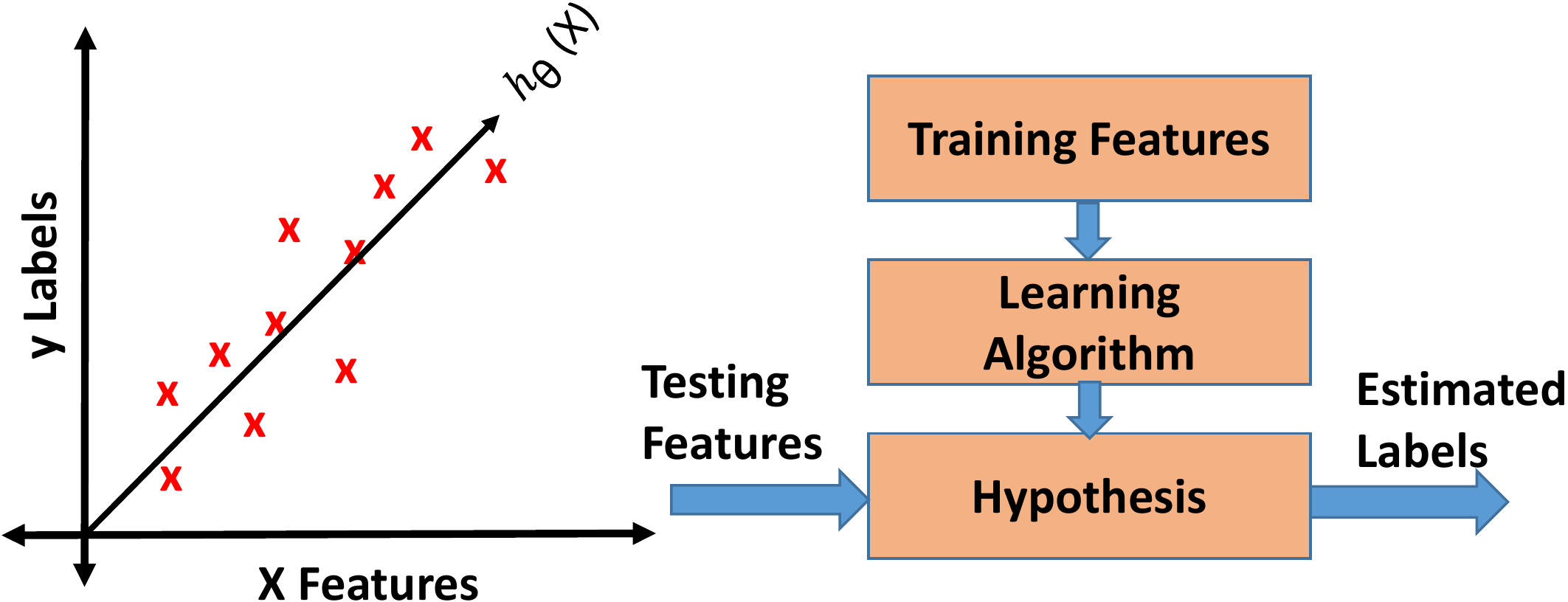}
	\caption{Single variable linear regression example.}
	\label{fig:figure9.pdf}
\end{figure}

In this work, we used linear regression as our supervised learning approach to build the prediction model. Learning regression is widely used in applications that require continuous output results. The process of building a prediction model included two stages, i.e. training and testing. The training of the model is based on finding the linear regression parameters required for mapping the input data (features) into output values (labels). The mapping function is denoted as the hypothesis. The features in our analysis were the independent variables of the weather conditions, while the labels are the dependent variables of the energy consumption. Fig. \ref{fig:figure9.pdf} shows an example of a single variable linear regression model. We denote the features matrix by $X = \{X^1, X^2, ..., X^n \}$ and the labels vector by $y = \{y^1, y^2, ..., y^n\}$, where $n$ is the number of input examples and each entry in $X$ represents the $i^{th}$ training example that could contain multiple features. The hypothesis $h_\theta(X^i)$ contains the parameters $\boldsymbol{\theta} = \{\theta_0, \theta_1, ..., \theta_m\}$ corresponds to the $i_{th}$ training example with $m$ features. Therefore, the hypothesis is given
\begin{equation}
\label{equation1}
h_\theta(X^i) = \theta_0 + \theta_1 X^i_1 + ... + \theta_m X^i_m,
\end{equation}
where we add a unity column to the features matrix corresponding the first values (bias parameters) in the hypothesis parameters. The goal is to minimize the cost function $J\{\theta_0,\theta_1, ... ,\theta_m\}$ using the mean squared error given by:
\begin{equation}
\label{equation4}
\underset{\boldsymbol{\theta}}{\mathrm{minimize}} \;\; J\{\theta_0,\theta_1,...,\theta_m\} = \frac{1}{2n}  \sum_{i = 1}^{n} (h_\theta(X^i) - y^i)^2,
\end{equation}
where $\boldsymbol{\theta}$ is calculated using the normal equation method
\begin{equation}
\label{theta}
 \boldsymbol{\theta} = (X^T X)^{-1} {X^T y},
\end{equation}
where $^T$ and $^{-1}$ are the matrix transpose and the inverse operations, respectively. The normal equation method is used to calculate the parameters of the hypothesis that minimize the cost function $J\{\theta_0,\theta_1,...,\theta_m\}$ as the number of features is small and it performs faster than the gradient descent method.
\subsection{Backward Elimination}
Generally, multi-variable linear regression applications include independent variables that are of more or less importance than others. For this reason we had to exclude the independent variables that were of less significance (lower significance level - SL) than the others. Backward elimination assisted in finding the most significant variables and pointed out the data necessary to fetch and thus reducing the amount of data being processed in the future forecast processes. Algorithm \ref{backeliminatoion} illustrates the procedure of the exclusion of less significant variables from our multiple regression model. The P-value is a correlation metric that gives us the significance level of each independent variable included in the regression model. In each loop of the backward elimination procedure we exclude the independent variable that has the highest P-value which has less significance for our model. There is a limit to compare that value and in our case we selected a value of 0.05.

\begin{algorithm}[h!]
	\caption{\\ Backward Elimination}
	\label{backeliminatoion}
	\begin{algorithmic}[1]
		\BState Calculate $P$ values for all features. 
		\BState Select a significant level ($SL$) to stay in the model.
		\BState Fit the regression model with all features.
		\While{$P \ge SL$} 
		\State Remove the feature with highest $P$ value from the feature set.
		\State Fit the regression model with the updated feature set.
		\State Recalculate the $P$ values.
		\EndWhile
		\BState Set of independent variables for the
		forecast model is ready.
	\end{algorithmic}
\end{algorithm}

\begin{table*}[t]
	\captionsetup{font=scriptsize}	
	\centering
	\caption{Coefficient of Determination for Single Independent Variable}
	\label{tab:single}
	\begin{tabular}{ l || l | l | l | l | l } \hline		
		\toprule
		\textbf{r-squared} & Atmospheric Pressure & Dew Point Temperature & Rainfall & Relative Humidity & Wind Chill \\ \hline
		\midrule		
		Active Power & 0.000 & 0.469 & 0.000 & 0.000 & 0.519  	
		\\  \hline
		Reactive power & 0.012 & 0.653 & 0.002 & 0.004 & 0.667 	
		\\ 
		\bottomrule    
	\end{tabular}
				\vspace{-0.2cm}	
\end{table*}

\begin{table*}[t]
	\captionsetup{font=scriptsize}
	\centering
	\caption{Coefficient of Determination for Multiple Independent Variables}
	\label{tab:multi}
	 \begin{threeparttable}
	\begin{tabular}{ l || l | l | l | l } \hline		
		\toprule
		\textbf{R-squared} & DPT - WC		
		  & DPT  - WC - AP & DPT - WC - AP - RH
		  & DPT -WC-AP-RH-RF \\ \hline
		\midrule		
		Active Power & 0.532 & 0.539 & 0.549 & 0.550	
		\\  \hline
		Reactive power & 0.704 & 0.705 &0.713 &0.713 	
		\\ 
		\bottomrule    
	\end{tabular}
		\begin{tablenotes}
		\scriptsize \item[*] DPT: Dew Point Temperature. WC: Wind Chill. AP: Atmospheric Pressure. RH: Relative Humidity. RF: Rainfall.
		\end{tablenotes}
	  \end{threeparttable}
	  				\vspace{-0.3cm}	  
\end{table*}

\begin{figure*}[t!]
	\centering
	\begin{multicols}{2}
		\includegraphics[width=8cm,height=5cm]{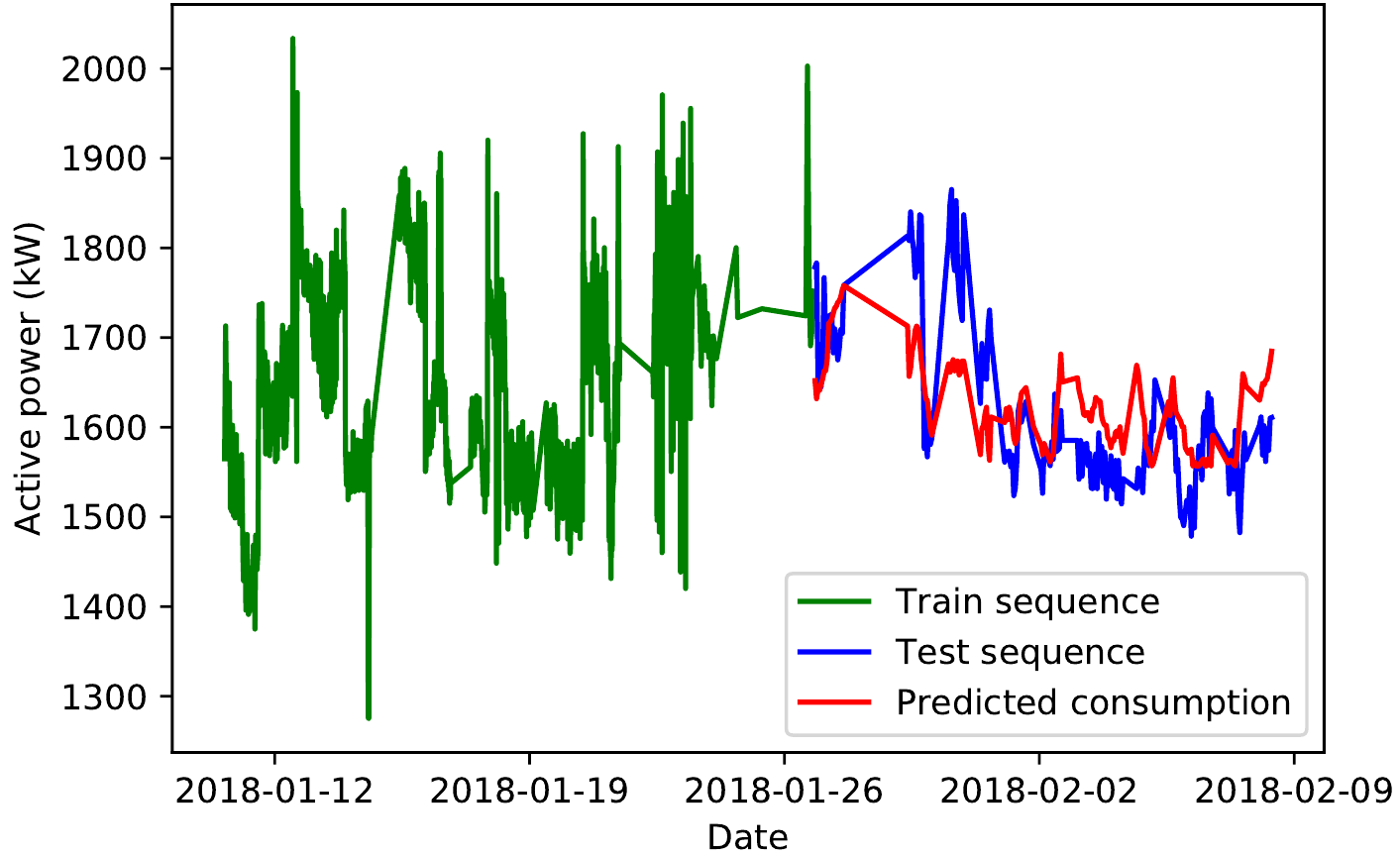}\caption{Active power forecast based on wind chill (high correlation). }\label{fig:result1.pdf}\par
		\includegraphics[width=8cm,height=5cm]{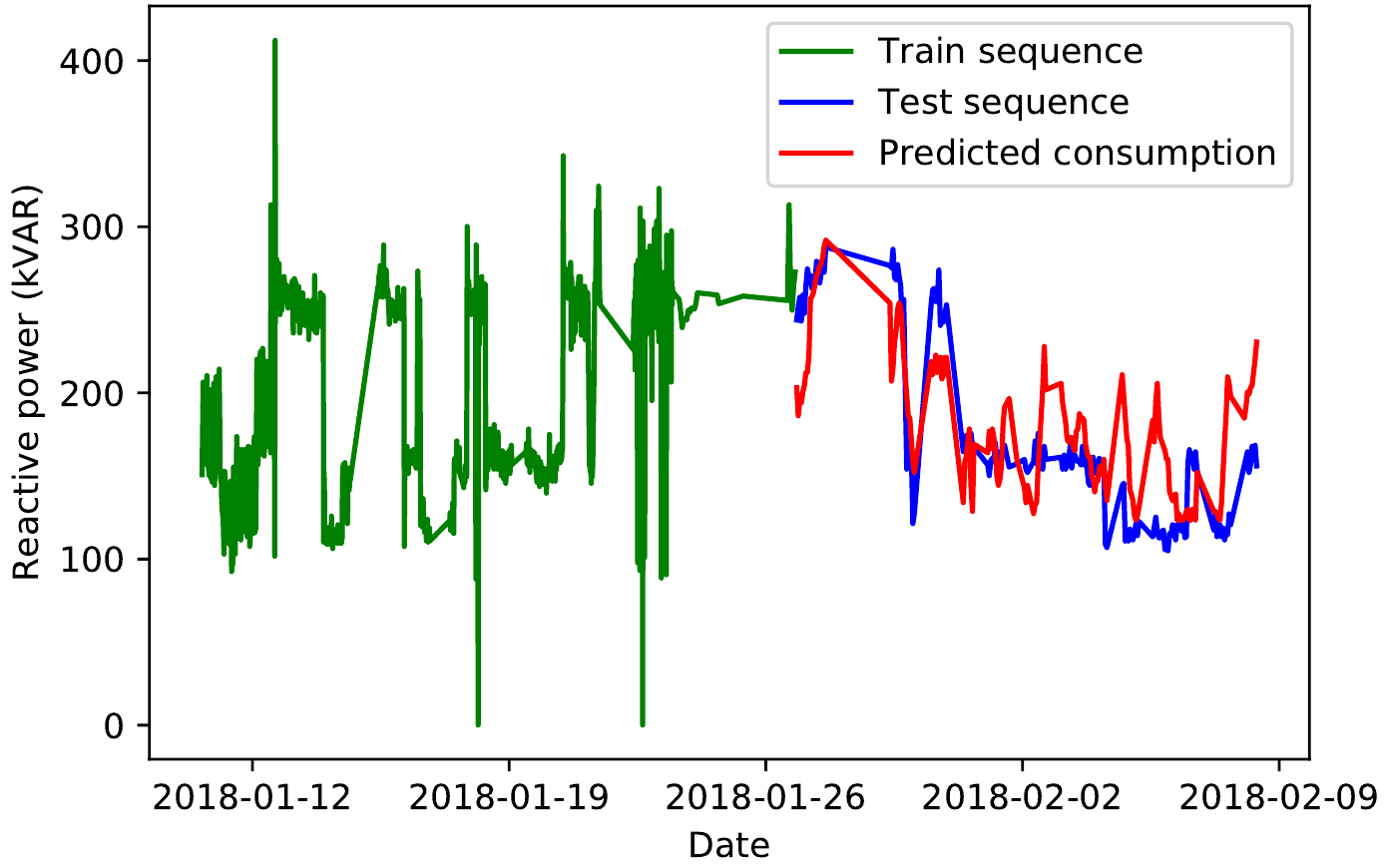}\caption{Reactive power forecast based on wind chill (high correlation). }\label{fig:result2.pdf}\par				
	\end{multicols}
			\vspace{-0.8cm}
\end{figure*}

\begin{figure*}[t!]
	\centering
	\begin{multicols}{2}
		\includegraphics[width=8cm,height=5cm]{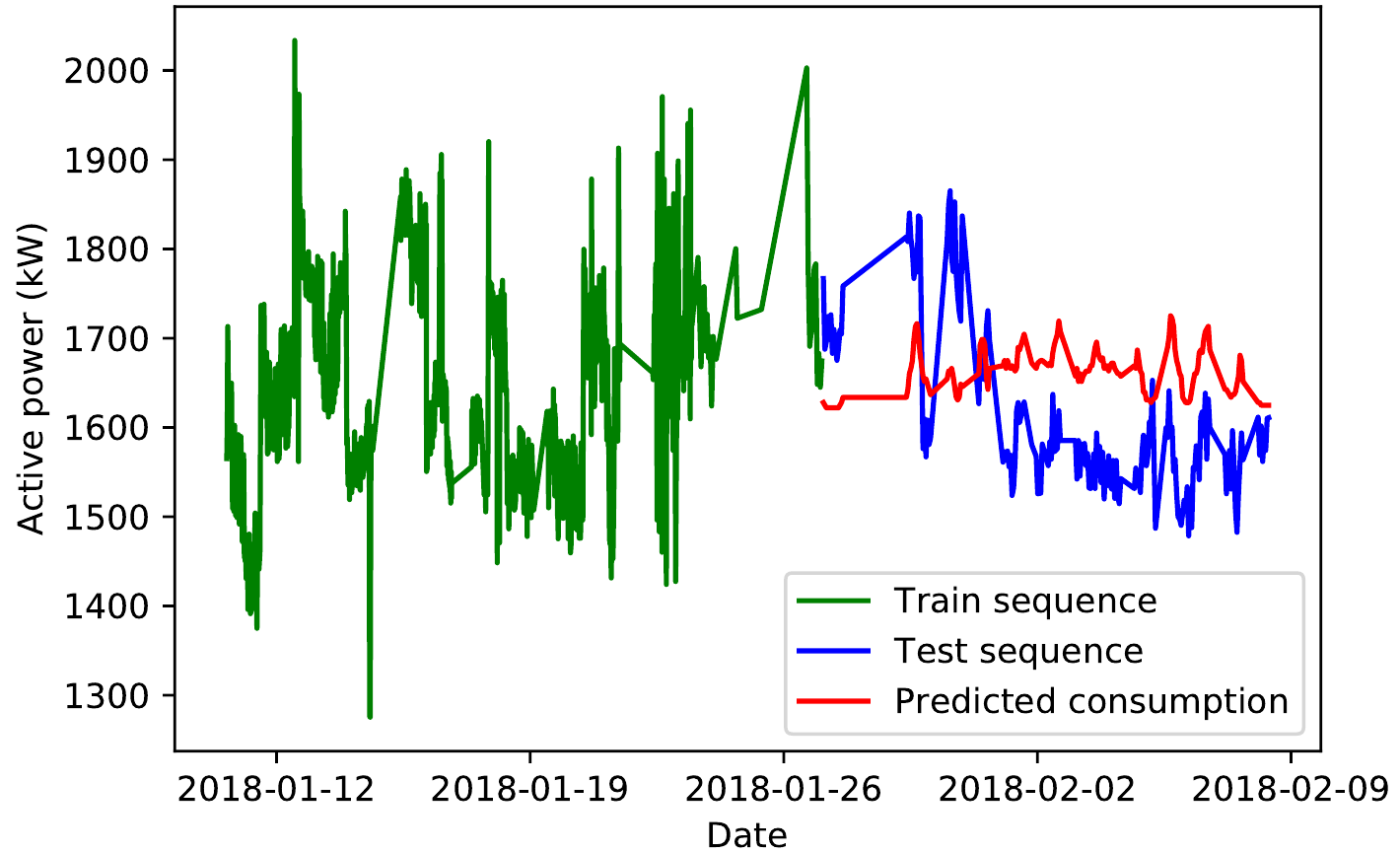}\caption{Active power forecast based on relative humidity (low correlation).}\label{fig:result3.pdf}\par		
		\includegraphics[width=8cm,height=5cm]{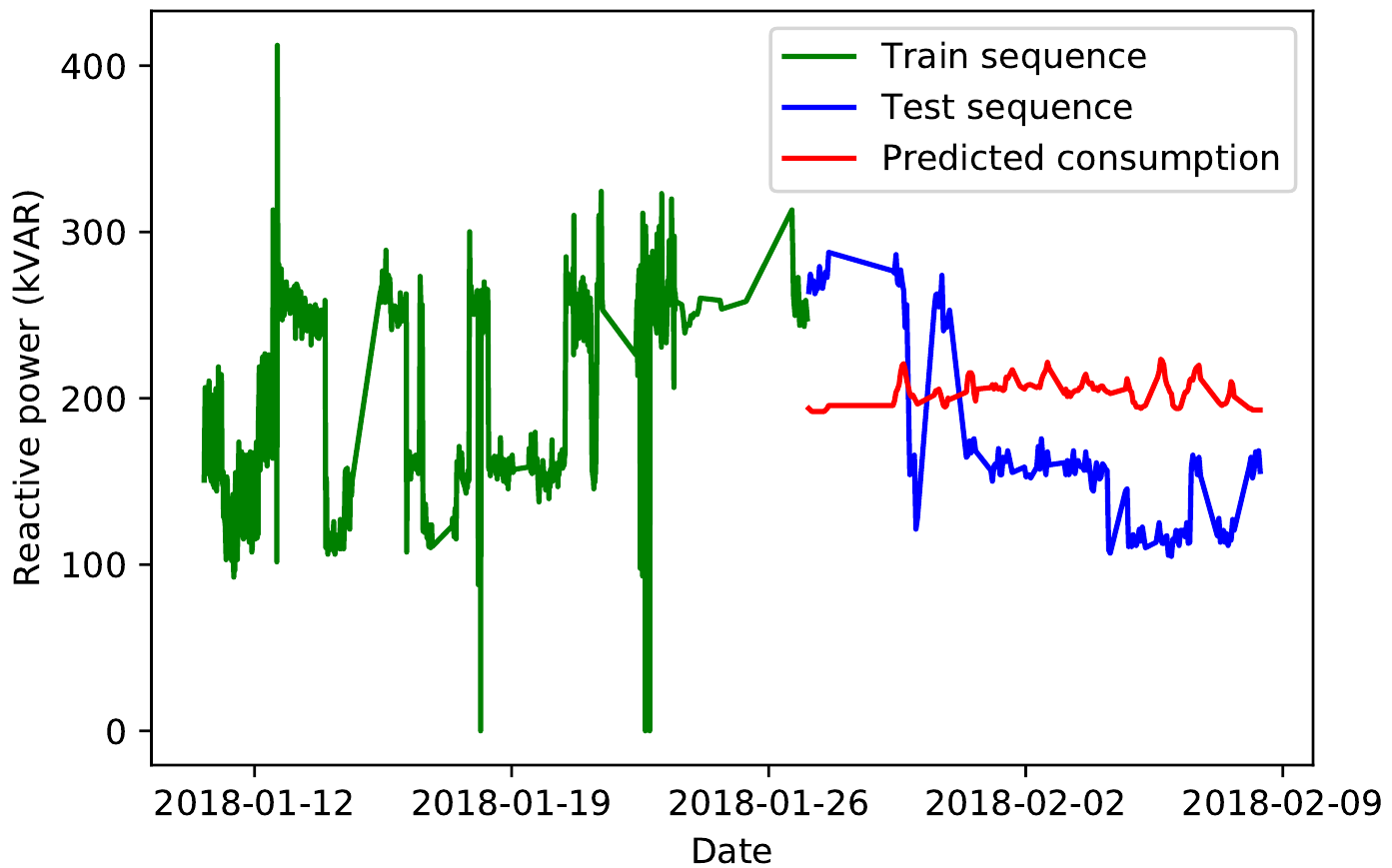}\caption{Reactive power forecast based on relative humidity (low correlation).}\label{fig:result4.pdf}\par			
	\end{multicols} 
			\vspace{-0.8cm}
\end{figure*}

\subsection{Model Building and Testing}
After the backward elimination stage, we focused on the sensor data that were more correlated and performed linear regression using \textit{linear regression libraries} in python to construct a regressor, i.e. classifier, that could fit the independent variables into the model and produce the power consumption forecast profile. The data from the csv files that were previously generated were imported to the main forecast algorithm and a linear regression model was generated. 
 
After the model building process, it was tested using a test sequence of data to compare the predicted values of active/reactive power and the output of the regression model. Additionally, due to the nature of the values that were observed (active/reactive power levels) we investigated the correlation of the weather conditions with their respective future active/reactive power values for the total number of sensors. The analysis of the time effects were undertaken offline to investigate the time delay response of the cooling system (cooling with air flows coming from the outside). More specifically, there could be some correlation between the near future (time shift correlation) power measurements and the current weather conditions.

\subsection{Energy Consumption Forecasting}
After feeding the training data to the prediction model, we were able to forecast the future energy consumption based on weather data. Afterwards, the forecast model was verified by using test data, i.e. the predicted consumption was compared with the actual consumed energy (test sequence). The percentage of training data were 80\% of the total available sensor data, and the rest was used as a test sequence for the comparison of the forecast with the real values.

\section{Experiment Results and Analysis}

In this section we present our experiment results and provide detailed analysis on each case. All the values of the weather conditions represent measurements obtained from the RealDC testbed through a period of almost two months. It is worth mentioning here that accurate forecasting models based on weather conditions require measurements obtained through a longer period of time to reflect the weather conditions at different seasons of the year. We leave this as part of our future targets where we plan to apply weather conditions of one year on our energy consumption forecasting model. We refer to all the measurements, either the active or reactive power, as \textit{historical} and \textit{future} values. The historical values indicate those measurements sensed by the testbed sensors prior to the prediction phase. On the other hand, the future values indicate \textit{actual} measurements sensed by the testbed sensors after the prediction phase. Also, the future values include the \textit{forecast} results that were obtained by the regression model. Tables \ref{tab:single} and \ref{tab:multi} illustrate the effectiveness of the independent variables, when used separately and when used as a combination, in predicting the energy consumption of the DC based on the coefficient of determination (CoD). The CoD is defined as the proportion of the variance in the dependent variable that is predictable from the independent variable(s). It is clear that combining multiple independent variables results in higher correlation and predictability of the dependent variables

From the results shown above, Fig. \ref{fig:result1.pdf} to Fig. \ref{fig:result4.pdf}, we can see that the power measurements vary a lot through the observed period of time. The reason for this is that the merged data are not continues and that there are gaps between active and reactive power measurements after data cleaning. The forecast results using single variable regression model shown in red color follow the same trend of the actual values. More explicitly, the energy consumption forecast of DCs can be obtained from forecast of the weather conditions due to the fact that weather conditions have a direct impact on the cooling system of DCs which has a large share of their energy consumption. The regression algorithm used the historical values of the weather conditions to calculate the parameters of the learning model based on the normal equations method and backward elimination. We have verified our forecast results by comparing them with the sensed measurements from the testbed over a period of two weeks. The verification was important to insure that reliable forecast values are produced by the regression model.

\begin{figure*}[t!]
	\centering
	\begin{multicols}{2}
		\includegraphics[width=8cm,height=5cm]{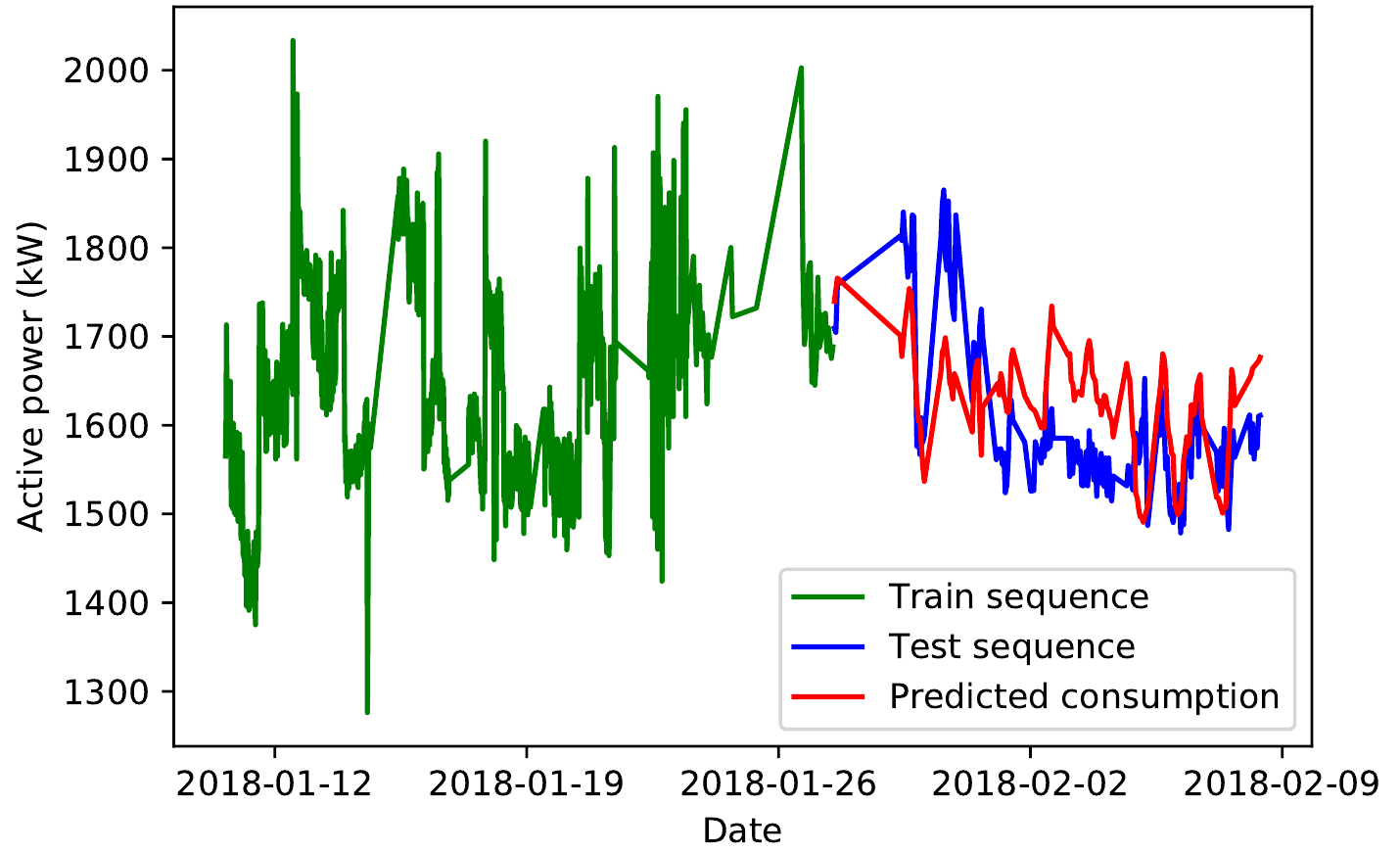}\caption{Active power forecast with multiple features of the weather conditions.}\label{fig:result5.pdf}\par
		\includegraphics[width=8cm,height=5cm]{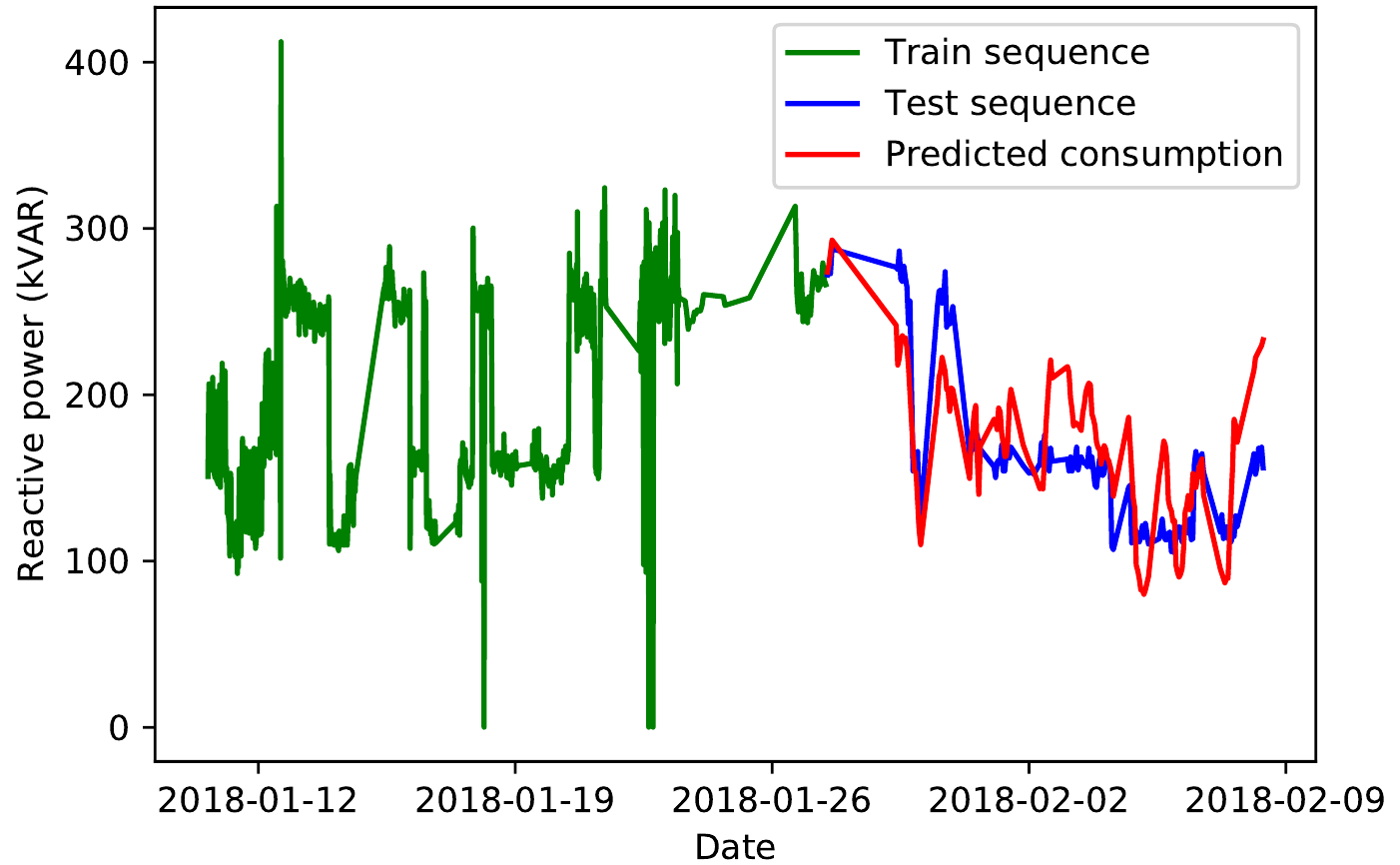}\caption{Reactive power forecast with multiple features of the weather conditions.}\label{fig:result6.pdf}\par			
	\end{multicols} 
	\vspace{-0.8cm}
\end{figure*}

Fig. \ref{fig:result1.pdf} and Fig. \ref{fig:result2.pdf} show the correlation between the energy consumption and the wind chill for the testbed based on both the active power and the reactive power. The wind chill was selected as it produced the highest statistical correlation following the procedure of feature selection that we explained earlier in the paper. It is important to note here that the RealDC testbed is based on sensor deployments at different locations in the data center of interest. Thus, the high correlation of either the active or the reactive power with a particular feature of the weather conditions reflects the fact that each sensor can provide different values depending on its location in the data center. For the sake of comparison, we show in Fig. \ref{fig:result3.pdf} and Fig. \ref{fig:result4.pdf} the forecast performance using low correlated features. It is clear that the forecast of the energy consumption is not accurate when using weather conditions with low correlation.

Finally, Fig. \ref{fig:result5.pdf} and Fig. \ref{fig:result6.pdf} show the energy consumption forecast of the active and the reactive power based on multiple features of the weather conditions. We have used atmospheric pressure, wind chill / dew point temperature and relative humidity as input features as they have the highest correlation. We observe that multiple features do not improve the forecast results when they are not highly correlated with the energy consumption.

\balance
\section{Conclusions and Future Work}
In this work we studied the correlation between the weather conditions and the energy consumption of data centers. Using real data obtained from the sensors of the RealDC testbed, part of the FIESTA-IoT platform, we investigated the correlation between the variations in weather conditions and how they affect the energy consumption. Our analysis showed that only certain weather features have significant impact on the energy consumption. We then used the correlated data to build a forecast model using linear regression algorithm. The experimental results showed that the forecast model of the energy consumption manages to make predictions based on the weather conditions with adequate accuracy. These results are indicative as they could provide DC operators and power distribution companies with tools to manage their power requirements. Our future work will include weather data of longer periods of time to provide more accurate forecast of the energy consumption. Larger training sets will enable us use more advanced machine learning techniques like artificial neural networks (ANN) and support vector machines (SVM).

\section*{Acknowledgment}
%{\fontsize{6.7pt}{}\selectfont
Part of this work has been funded by the FIESTA-IoT project, with GA number: 643943. The REAL-DC testbed has been utilized for gathering the data that have been used in this work. Georgios Smpokos is a researcher at CYTA-Hellas and funded by the WiVi-2020 project. Mohamed Elshatshat contributed to this work during his secondment to CYTA-Hellas from ICS-FORTH and funded by the WiVi-2020 project. The WiVi-2020 project has received funding from the European Unions Seventh Framework Programme (FP7/2007-2013) under grant
agreement no 324515 and from the European Unions Horizon
2020 research and innovation programme under the Marie
Sklodowska-Curie grant agreement No 642743.

\bibliographystyle{IEEEtran}
{\footnotesize \bibliography{Ref}}


\end{document}